# Negative magneto-resistance of electron gas in a quantum well with parabolic potential

F. M. Hashimzade, Kh. A. Hasanov and M. M. Babayev

*Institute of Physics, Academy of Sciences of Azerbaijan, AZ 1143, Baku, Azerbaijan*

We have studied the electrical conductivity of the electron gas in parallel electric and magnetic fields directed along the plane of a parabolic quantum well (across the profile of the potential). We found a general expression for the electrical conductivity applicable for any magnitudes of the magnetic field and the degree of degeneration of the electron gas. A new mechanism of generation of the negative magneto-resistance has been revealed. It has been shown that in a parabolic quantum well with a non-degenerated electron gas the negative magneto-resistance results from spin splitting of the levels of the size quantization.

## 1. INTRODUCTION

Since recently a great deal of attention is being given to study of the transport phenomena in systems the size of which is comparable with the de Broglie wave length of electrons. The confinement of motion of electrons in such systems results in the transport phenomena being sharply different from transport phenomena in large samples. An interesting issue is also the effect of a magnetic field on the conductivity of electron gas in a quantum well.

The change of resistance in a magnetic field (magneto-resistance) may occur due to the effect of two factors: firstly, because of the change in concentration of the carriers of current in the magnetic field; secondly, because of the change in mobility of the carriers. The second factor of the occurrence of the magneto-resistance is of a special interest as it provides important information about the effective mass of an electron, the scattering mechanism and the parameters of the quantum well.

There is a number of studies of magneto-resistance of electron gas in quantum wells [1-7]. In papers [2, 4-6] the main attention was given to a quantum well with an infinite rectangle potential. In papers [1, 3, 7] the magneto-resistance of a quantum well with parabolic potential was studied. However, in these papers the effect of electron spin was not taken into account. Below we shall see that accounting for the latter factor results in a qualitatively new result.

In the present paper we have studied the dependence of electrical conductivity of electron gas in a quantum well with parabolic potential on the magnetic field parallel to the electric field, directed along the plane of the quantum well (across the profile of the potential), taking into account the electron spin. It has been assumed that the concentration of electrons does not depend on magnetic field. An expression has been obtained applicable for any degree of degeneration of electron gas and magnetic field. We found that in a quantum well with a non-degenerated electron gas the magneto-resistance in a certain range of magnetic field is negative. The study of the mechanism of the negative magneto-resistance (NMR) reveals that here the decisive role is played by spin splitting of oscillator levels (NMR does not occur if spin is neglected).

We should note that NMR has been long observed in the systems with three-dimensional, as well as with two-dimensional and one-dimensional electron gas. It occurs in transverse, as well as in longitudinal magnetic field in a certain temperature range. Such diversity in the conditions under which the NMR occurs suggests that its mechanisms are different [8].

In paper [9] the magneto-resistance of a three-dimensional crystal in a longitudinal $\left(\vec{B}\|\vec{E}\right)$ quantizing magnetic field has been calculated, and it has been shown that in the range of magnetic field satisfying the condition $\hbar\omega_c \leq \kappa_0 T$ (where $\omega_c$ is the cyclotron frequency of the electrons, and $T$ is the temperature) the NMR occurs in non-degenerated semiconductors. In this case the NMR arises from a specific, jump-like dependence of the density of states of electrons in a quantizing magnetic field. In other words, it occurs because in the range $\hbar\omega_c \leq \kappa_0 T$ the probability of scattering is smaller than the one without magnetic field. In this mechanism the spin of electrons plays no significant role (in [9] the spin was not taken into account).

In paper [8] the magneto-resistance was studied for the case of electrons scattering on a deformation potential of a three-dimensional crystal in a transverse $\left(\vec{B}\perp\vec{E}\right)$ magnetic field. It has been shown that under a consistent account of quantization in a magnetic field the NMR occurs in transverse magnetic fields as well.

In explaining the occurrence of the NMR in both three- and two-dimensional systems the major basis in the theory of weak localization of electrons [10-12] and the quantum corrections to the conductivity as a result of the latter. This theory can be equally applied to the NMR occurring in longitudinal as well as in transverse magnetic field [4, 13-19]. In this mechanism the Landau quantization and the size quantization plays no significant role: the effect arises in weak (classical) magnetic fields as well.

In [6] the magneto-resistance was studied in a magnetic field directed along the plane of a layer of a strongly degenerated two-dimensional electron gas with a width close to zero. It has been shown that for electronic scattering on the ions of admixture the NMR occurs because the magnetic field causes the spin polarization, which changes the screening properties. In this paper no account was taken of the effect of the magnetic field on the energy spectrum of the electrons.

We present the study of the case where the NMR in a quantum well with parabolic potential arises from the effect of electron spin on the energy spectrum. Here the NMR occurs under relatively low magnitudes of the lattice temperature and the magnetic field. Under the higher magnitudes of the magnetic field the magneto- resistance of a non-degenerated electron gas becomes positive and grows monotonically.

## 2. SPECTRUM AND DENSITY OF STATES OF ELECTRONS IN A QUANTUM WELL

### 2.1. Model of a quantum well

Here we consider quantum wells based on $A^3 B^5$ semiconductors. A typical example is a $GaAs/Al_x Ga_{1-x} As$ quantum well which has been most extensively studied experimentally. In relatively wide quantum wells ($L_x \geq 100\, nm$) the potential can be created of parabolic form with depth $\Delta_1$ restricted by a barrier of height $\Delta_2$ [3]. We are considering the case where the average energy of electrons, $\bar{\varepsilon}$, is small compared to the depth $\Delta_1$ of the parabola (for a degenerated electronic gas the Fermi energy satisfies the condition $\varepsilon_f < \Delta_1$, and for a non-degenerated gas - $k_0 T \ll \Delta_1$. In this case for the electrons in the quantum well the potential may be written as

$$U = \frac{m\omega_0^2 x^2}{2} \qquad (1)$$

(the x-axis is perpendicular to the layer of electronic gas). Here $m$ is the effective mass of electrons, $\omega_0$ is the parameter of the parabolic potential determined from

$$\Delta_1 = \frac{m\omega_0^2 L_x^2}{8}, \qquad (2)$$

where $L_x$ is the width of the quantum well.

### 2.2. Energy spectrum of electrons and wave function

Let the induction of the magnetic field lie in the y-z plane of the layer of electron gas parallel to z-axis. Choosing the Landau calibration for the vector potential as $A(0, x \cdot B, 0)$ one can present the law of dispersion of electrons as [20,21]:

$$\varepsilon_{N,k_y,k_z,\sigma} = \left(N + \frac{1}{2}\right)\hbar\omega + \frac{\hbar^2 k_z^2}{2m} + \frac{\omega_0^2}{\omega^2}\frac{\hbar^2 k_y^2}{2m} + \sigma g\mu_B B, \qquad (3)$$

where $\omega = \sqrt{\omega_o^2 + \omega_c^2}$, $\omega_c = \frac{eB}{mc}$ is the cyclotron frequency of electrons, $e$ is the charge of an electron, $c$ is the light speed, $g$ is the spectroscopic splitting factor, $\sigma = \pm\frac{1}{2}$ is the spin quantum number, $N$ is the oscillation quantum number, $\mu_B = \frac{e\hbar}{2m_0 c}$ is Bohr magneton, $m_0$ is the mass of free electrons.

The electronic wave functions corresponding to the eigenvalues of (3) have the following form:

$$\varphi_{N,k_y,k_z}(r) = \varphi_N(x - x_0) e^{i(k_y y + k_z z)}, \qquad (4)$$

where

$$\varphi_N(x - x_0) = \frac{1}{\pi^{1/4}\alpha^{1/2}(2^N N!)^{1/2}} \exp\left[-\frac{(x-x_0)^2}{2\alpha^2}\right] H_N\left(\frac{x-x_0}{\alpha}\right) \qquad (5)$$

is the wave function of a harmonic oscillator with the centre at $x_0$ and the quantum number $N$, and $H_N(x)$ is the Hermite polynomial [22],

$$\alpha = \sqrt{\frac{\hbar}{m\omega}}, \quad x_0 = \frac{\omega_c}{\omega}\alpha^2 k_y. \qquad (6)$$

### 2.3. Density of states of electrons

The general expression for the density of states is

$$g(\varepsilon) = \sum_{N,\sigma,k_y,k_z} \delta(\varepsilon_{N,\sigma,k_y,k_z} - \varepsilon). \qquad (7)$$

Replacing summation over $k_y$ and $k_z$ by integration over $x_0$ and $\varepsilon$, using (3) and (6), we obtain for the density of states of electrons in a quantum well [23]:

$$g(\varepsilon) = \frac{L_y L_z m}{\pi^2 \hbar^2}\frac{\omega}{\omega_0}\sum_{N,\sigma} I(N,\sigma,\varepsilon), \qquad (8)$$

where $L_y$ and $L_z$ are the lengths along $y$ and $z$ directions, and the following notations are used:

$$I(N,\sigma,\varepsilon) = \int_0^{x_{0m}} \frac{dx_0}{\sqrt{\frac{\varepsilon - \varepsilon_{N,\sigma}}{\beta} - x_0^2}}, \qquad (9)$$

$$\varepsilon_{N\sigma} = \left(N + \frac{1}{2}\right)\hbar\omega + \sigma g \mu_B B \,, \quad \beta = \frac{m\omega_0^2 \omega^2}{2\omega_c^2}. \tag{10}$$

In general, the upper limit $x_{0m}$ of integration in (9) depends on the parameters $N, \sigma, \varepsilon$ and $L_x$. To determine $x_{0m}$ we take into account the fact that the summation in the (8) and (9) is taken for the positive values of the expression under the square root and over the oscillators with the centres in the range $|x_0| < \frac{L_x}{2}$. Here, two cases are possible:

1. $\varepsilon - \varepsilon_{N,\sigma} < \beta\left(\frac{L_x}{2}\right)^2$. In this case the centres of all oscillators corresponding to the positive value of the expression under the square root are within the range $|x_0| < \frac{L_x}{2}$ and, therefore, $x_{0m} = \sqrt{(\varepsilon - \varepsilon_{N,\sigma})/\beta}$. As it can be seen from (9), $I(N, \sigma, \varepsilon)$ does not depend on $\varepsilon$ and $L_x$:

$$I(N, \sigma, \varepsilon) = \frac{\pi}{2}. \tag{11}$$

2. $\varepsilon - \varepsilon_{N,\sigma} > \beta\left(\frac{L_x}{2}\right)^2$. In this case $x_{0m} = \frac{L_x}{2}$ and $I(N, \sigma, \varepsilon)$ depends on $\varepsilon, N, \sigma$ and $L_x$:

$$I(N, \sigma, \varepsilon) = \arcsin \frac{L_x \sqrt{\beta}}{2\sqrt{\varepsilon - \varepsilon_{N,\sigma}}}. \tag{12}$$

Using (11) and (12) in (8) we obtain the following expression for the density of states:

$$g(\varepsilon) = \frac{L_y L_z}{2\pi\hbar} \frac{m\omega}{\omega_0} \left\{ \sum_{N,\sigma} H\left(-\varepsilon + \varepsilon_{N,\sigma} + \frac{\beta L_x^2}{2}\right) H(\varepsilon - \varepsilon_{N,\sigma}) + \right.$$

$$\left. + \frac{2}{\pi} \sum_{N,\sigma} H\left(\varepsilon - \varepsilon_{N,\sigma} - \frac{\beta L_x^2}{2}\right) \arcsin \frac{L_x \sqrt{\beta}}{2\sqrt{\varepsilon - \varepsilon_{N,\sigma}}} \right\}, \tag{13}$$

where $H(x)$ is Heaviside function,

$$H(x) = \begin{cases} 1, & x \geq 0; \\ 0, & x < 0. \end{cases} \tag{14}$$

In the most of experimentally studies cases the average energy of electrons ($\bar{\varepsilon}$) satisfies the conditions $\bar{\varepsilon} - \frac{\hbar\omega}{2} - \frac{1}{2}|g|\mu_B B < \Delta_1 < \frac{\beta L_x^2}{2}$. As mentioned above, we consider this particular case in the study of the

electrical conductivity. Then, in (13) for the density of states of electrons one can neglect the second term, and $g(\varepsilon)$ becomes a step function of $\varepsilon$:

$$g(\varepsilon) = \frac{L_y L_z m}{2\pi\hbar^2} \cdot \frac{\omega}{\omega_0} \sum_{N\sigma} H(\varepsilon - \varepsilon_{N\sigma}). \quad (15)$$

For every given value of $\varepsilon$, $g(\varepsilon)$ is a saw-tooth function of the magnetic field.

## 3. CHEMICAL POTENTIAL OF ELECTRONS IN A QUANTUM WELL

Taking the bottom of the quantum well as the origin, the surface density of the two-dimensional electron gas can be written as

$$n = \frac{1}{L_y L_z} \int_0^{\varepsilon} g(\varepsilon) f_0(\varepsilon) d\varepsilon, \quad (16)$$

where

$$f_0(\varepsilon) = \left[1 + \exp\left(\frac{\varepsilon - \xi}{k_0 T}\right)\right]^{-1} \quad (17)$$

is the distribution function of the electrons in equilibrium (Fermi-Dirac distribution), and $\xi$ is the chemical potential of the electrons.

After substitution of (15) and (17) into (16) and integration over the energy we obtain the following expression for the surface density of electrons:

$$n = \frac{n_0}{2} \frac{\omega}{\omega_0} \sum_N \left\{ \ln\left[e^{\eta - \left(N + \frac{1}{2}\right)a + \frac{b}{2}} + 1\right] + \ln\left[e^{\eta - \left(N + \frac{1}{2}\right)a - \frac{b}{2}} + 1\right] \right\}, \quad (18)$$

where

$$n_0 = \frac{m k_0 T}{\pi \hbar^2}, \quad a = \frac{\hbar \omega}{k_0 T}, \quad b = \frac{|g| \mu_B B}{k_0 T}, \quad \eta = \frac{\xi}{k_0 T}. \quad (19)$$

The chemical potential can be found from (18) for an arbitrary degree of degeneration of electron gas and arbitrary magnetic field. Here we assume that the surface density of electrons does not depend on the magnetic field. Below we investigate how the chemical potential depends on $n$ and $B$ in the cases of non-degenerated and strongly degenerated electronic gas.

### 3.1. Non-degenerated electron gas

The electron gas in a quantum well is non-degenerated $(f_0(\overline{\varepsilon}) \ll 1)$ if the following condition holds:

$$-\eta_0 \equiv \frac{a}{2} - \frac{b}{2} - \eta \gg 1. \qquad (20)$$

Under this condition from (18) we obtain for the chemical potential

$$\xi(B) = k_0 T \ln\left[\frac{2 sh\frac{a}{2}}{ch\frac{b}{2}} \frac{n}{n_0} \frac{\omega_0}{\omega}\right]. \qquad (21)$$

For a given value of the surface density one can find from (20) and (21) the temperature range that satisfies the non-degeneracy condition. For example, for $GaAs$ ($m = 0,067 m_0$, $\hbar\omega = 2,91 meV$ [25]) the non-degeneracy condition ($\eta_0 \leq -3$) is satisfied at $T \geq 1K$ for $n = 10^8\ cm^{-2}$, at $T \geq 8K$ - for $n = 10^9\ cm^{-2}$, and at $T \geq 60K$ - for $n = 10^{10}\ cm^{-2}$.

In the limiting case $a \gg 1$, $b \gg 1$ the expression (21) can be written as

$$\xi(B) = \frac{\hbar\omega}{2} - \frac{1}{2}|g|\mu_B B + k_0 T \ln\left(\frac{2n}{n_0}\frac{\omega_0}{\omega}\right), \qquad (22)$$

and in the limiting case $a \gg 1$, $b \ll 1$ it can be written as

$$\xi(B) = \frac{\hbar\omega}{2} + k_0 T \ln\left(\frac{2n}{n_0}\frac{\omega_0}{\omega}\right). \qquad (23)$$

### 3.2. Strongly degenerated electron gas

For the degenerated electronic gas we have $\eta_0 \geq 0$. We are considering the case of a strongly degenerated electron gas:

$$\eta_0 \equiv \eta - \frac{a}{2} + \frac{b}{2} \gg 1. \qquad (24)$$

Under this condition we can omit 1 in the argument of the logarithmic function in (18) (this is equivalent to the replacement of the function $\left(-\frac{\partial f_0}{\partial \varepsilon}\right)$ by delta-function $\delta(\varepsilon - \xi)$). In this case the analytical expression for the function of the chemical potential depends on the energy range where it falls (see Fig. 1).

If the chemical potential is between the sublevels of the same oscillator level, i.e. if $\left(N + \frac{1}{2}\right)a - \frac{b}{2} < \eta < \left(N + \frac{1}{2}\right)a + \frac{b}{2}$ then form (18) we obtain

$$\xi(B) = \left[\frac{(N_0 - 1)^2}{4L} + \frac{1}{2}\right]\hbar\omega + \frac{2k_0 T}{L}\frac{n}{n_0}\frac{\omega_0}{\omega} - \frac{|g|\mu_B B}{2L} \qquad (25)$$

When the chemical potential falls between two neighbouring oscillator levels, i.e. if $\left(N+\frac{1}{2}\right)a+\frac{b}{2}<\eta<\left(N+\frac{3}{2}\right)a-\frac{b}{2}$ we obtain

$$\xi(B) = \frac{L}{4}\hbar\omega + \frac{2k_0 T}{L}\frac{n}{n_0}\frac{\omega_0}{\omega}. \tag{26}$$

Here $L$ is the total level of sublevels below the chemical potential.

For a given value of the surface density of electrons the energy of the N-th oscillator level depends on the magnetic field as $\left(N+\frac{1}{2}\right)\hbar\omega_0\left(1+\frac{\omega_c^2}{\omega_0^2}\right)^{1/2}$, whereas $\xi(B)$ depends on the magnetic field only weakly (as the magnetic field grows the first terms in (25) and (26) are growing and the second terms are decreasing, and as a result they partially offset each other). Therefore, as the magnetic field grows, the oscillator levels move upwards, crossing the chemical potential in turn. Corresponding to this, for $\xi(B)$ expressions (25) or (26) are used alternatively. It is clear that for $\eta=\left(N_0+\frac{1}{2}\right)a-\frac{b}{2}$ and $\eta=\left(N_0+\frac{1}{2}\right)a+\frac{b}{2}$ expressions (25) and (26) must coincide. However, one must take into account that in the transition from (25) to (26) (at the transition magnitudes of energy) $L$ changes by one unit.

If the spin splitting is neglected $(b\to 0)$ one can derive the following expression for the chemical potential, applicable at any magnitude of the magnetic field:

$$\xi(B) = (N_0+1)\frac{\hbar\omega}{2} + \frac{k_0 T}{N_0+1}\cdot\frac{n}{n_0}\frac{\omega_0}{\omega}. \tag{27}$$

Here $N_0=\frac{L}{2}-1$ is the number of the highest oscillator level below the chemical potential. Because the g-factor for *GaAs* is small, the result obtained from (27) almost coincides with the result obtained from (25)-(26).

By definition,

$$\left(N_0+\frac{3}{2}\right)\hbar\omega > \xi(B) \geq \left(N_0+\frac{1}{2}\right)\hbar\omega. \tag{28}$$

Then from (27) and (28) we obtain the expression for $N_0$:

$$N_0 = \left[\sqrt{\frac{2\pi n\hbar\omega_0}{m\omega^2}+\frac{1}{4}}-\frac{1}{2}\right], \tag{29}$$

where [*x*] is the integer part of *x*.

For a given value of the **surface density** one can easily get from (27) the magnitude of the magnetic field at which the *N*-th oscillator level crosses the chemical potential:

$$B_N = \frac{m\omega_0}{e}\sqrt{\frac{2\pi n\hbar}{m\omega_0 N(N+1)} - 1} \ . \qquad (30)$$

Note that under strong degeneration the chemical potential lies above the 0-th oscillator level for any magnitude of the magnetic field, i.e. $N \geq 1$ in (30). At the values of the surface density $n < \frac{m\omega_0}{\pi\hbar}$ the chemical potential is below the first oscillator level even without magnetic field; the electrons are mostly at the zero oscillator level. In this case the expression for the chemical potential is

$$\xi(B) = \frac{\hbar\omega}{2} + \frac{\pi\hbar^2}{m} n \frac{\omega_0}{\omega} . \qquad (31)$$

For the strongly degenerated electron gas $(T \to 0)$ the condition of the electrons locating at the zero oscillator level for *GaAs* ($m = 0.067 m_0$, $\hbar\omega_0 = 2.91\,meV$ [25]) is $n < 9\cdot 10^{10}\,cm^{-2}$, and for *InSb* ($m = 0.016 m_0$, $\hbar\omega_0 = 7.5\,meV$ [24]) the condition is $n < 5\cdot 10^{10}\,cm^{-2}$. Note that in *InSb* ($g = -51,2$) because of the large spin splitting of the oscillator levels in the magnetic field even the upper spin sublevel of the zero level moves upwards crossing the chemical potential (at the density $n = 5\cdot 10^{10}\,cm^{-2}$ this occurs in a magnetic field with $B > 2{,}2\,T$ ), and the electrons are located only at the lower spin sublevel.

## 4. CALCULATION OF ELECTRICAL CONDUCTIVITY

Consider an electron gas in a quantum well in external electric ($\vec{E}$) and longitudinal magnetic ($\vec{B}$) fields: $Ox \perp \vec{E} \parallel \vec{B} \parallel Oz$. Because the magnetic field does not affect the movement of an electron along z-axis, one can use Boltzmann transport equation for the distribution function $f(N, \sigma, k_y, k_z)$ of the electrons [26]. Then for the density of current one can use the following:

$$j_z = \sum_{N, k_y, k_z, \sigma} (-e) f(N, \sigma, k_y, k_z) \upsilon_z , \qquad (32)$$

where $(-e)$ is the charge and $\upsilon_z = \frac{\hbar k_z}{m}$ is the velocity of an electron. Using the relaxation time approximation we can write [26]:

$$f(N, \sigma, k_y, k_z) = f_0(\varepsilon) + \upsilon_z \tau_B(\varepsilon) \frac{\partial f_0}{\partial \varepsilon} e E_z , \qquad (33)$$

where $\tau_B(\varepsilon)$ is the relaxation time of electrons in the magnetic field. We are considering the following electronic scattering mechanisms: scattering on acoustic phonons, on point defects (on a short-range potential of the form $(V(\vec{r}) = V_0 \delta(\vec{r}))$ and on alloy disorder. For all these cases the relaxation time can be written as [26, 27]:

$$\tau_B(\varepsilon) = \tau_1 \cdot g^{-1}(\varepsilon), \qquad (34)$$

where $g(\varepsilon)$ is the density of states of electrons in the magnetic field.

After substitution of (33) and (34) into (32), change of variable $k_y' = k_y \dfrac{\omega_0}{\omega}$ and integration over $(k_y', k_z)$ plane in polar coordinates we obtain:

$$j_z = \frac{e^2 L_y L_z}{2\pi\hbar^2} \cdot \frac{\omega}{\omega_0} \tau_1 E_z \sum_{N\sigma} \int_0^\infty (\varepsilon - \varepsilon_{N,\sigma}) g^{-1}(\varepsilon) \left(-\frac{\partial f_0}{\partial \varepsilon}\right) d\varepsilon. \qquad (35)$$

The density of states, $g(\varepsilon)$, in (35) for the general case is determined by (13). In most cases studied empirically for the density of states one can use (15). Then, replacing in (35) $\varepsilon - \varepsilon_{N,\sigma}$ by $x k_0 T$, for the electrical conductivity we obtain:

$$\sigma_{zz}(B) = -\frac{e^2 \tau_1}{m} k_0 T \sum_{N,\sigma} \int_0^\infty \frac{x \dfrac{d}{dx} f_0\left[x + \left(N + \dfrac{1}{2}\right)a - \sigma b\right]}{\sum_{N_1, \sigma_1} H[x + (N - N_1)a - (\sigma - \sigma_1)b]} dx. \qquad (36)$$

Dividing the range of integration in (36) as $\displaystyle\int_0^\infty \to \sum_{r=0}^\infty \int_{ra}^{(r+1)a}$ and integration by parts we obtain the following for $\sigma_{zz}(B)$:

$$\sigma_{zz}(B) = \frac{e^2 \tau_1}{2m} k_0 T \sum_{N=0}^\infty \left\{ \frac{N+1}{2N+1}\left[Na f_0\left(\left(N+\frac{1}{2}\right)a - \frac{b}{2}\right) - (Na+2b) f_0\left(\left(N+\frac{1}{2}\right)a + \frac{b}{2}\right) + \right.\right.$$

$$+ 2\ln\frac{1+e^{\eta-\left(N+\frac{1}{2}\right)a+\frac{b}{2}}}{1+e^{\eta-\left(N+\frac{1}{2}\right)a-\frac{b}{2}}} + (Na+b) f_0\left(\left(N+\frac{1}{2}\right)a + \frac{b}{2}\right) - (Na+2a-b) f_0\left(\left(N+\frac{3}{2}\right)a - \frac{b}{2}\right) +$$

$$+ 2\ln\frac{1+e^{\eta-\left(N+\frac{1}{2}\right)a-\frac{b}{2}}}{1+e^{\eta-\left(N+\frac{3}{2}\right)a+\frac{b}{2}}} + \frac{N+1}{2N+3}\left[((N+2)a - 2b) f_0\left(\left(N+\frac{3}{2}\right)a - \frac{b}{2}\right) - \right.$$

$$\left.\left. - (N+2)a f_0\left(\left(N+\frac{3}{2}\right)a + \frac{b}{2}\right) + 2\ln\frac{1+e^{\eta-\left(N+\frac{3}{2}\right)a+\frac{b}{2}}}{1+e^{\eta-\left(N+\frac{3}{2}\right)a-\frac{b}{2}}}\right]\right\}. \qquad (37)$$

This formula is applicable for arbitrary degree of degeneracy of electron gas. To analyze how the electrical conductivity depends on the magnetic field, the temperature and the parameters of the crystal let us consider a quantum well with a non-degenerated and a strongly degenerated electron gas.

### 4.1. Non-degenerated electron gas

If the condition (20) holds, it is possible to compute the sum over N in (37). Then from (21) and (37) we obtain the following expression for the electrical conductivity of a quantum well with a non-degenerated electron gas:

$$\sigma_{zz}(B) = \sigma(0)\frac{\omega_0}{\omega}\frac{sh\frac{a}{2}}{ch\frac{b}{2}}e^{-\frac{a}{2}}\left[(2-b)e^{\frac{b}{2}} + a\,sh\frac{b}{2} - \frac{a}{e^a-1}ch\frac{b}{2} - \left(\frac{a}{2}-b\right)e^{\frac{a}{2}}sh\frac{b}{2}\ln cth\frac{a}{4}\right]. \quad (38)$$

Here

$$\sigma(0) = \frac{e^2\tau_1}{m}\frac{\pi\hbar^2}{m}n \quad (39)$$

is the electrical conductivity without magnetic field $(B=0)$ in the temperature range $k_0 T \ll \hbar\omega_0$.

When $g \to 0$ from (38) we derive the dependence of the conductivity on the magnetic field:

$$\sigma_{zz}(B) = \sigma(0)\frac{\omega_0}{\omega}\left[1 - \left(1 + \frac{a_0}{2}\frac{\omega}{\omega_0}\right)e^{-a_0\frac{\omega}{\omega_0}}\right], \quad (40)$$

where

$$a_0 = \frac{\hbar\omega_0}{k_0 T}; \qquad \frac{\omega}{\omega_0} = \sqrt{1 + \left(\frac{e}{m\omega_0}\right)^2 B^2}. \quad (41)$$

It follows from (40) that neglecting the spin results in the conductivity being a monotonically decreasing function of the magnetic field $B$.

For the case $a \gg 1$ from (38) we obtain:

$$\sigma_{zz}(B) = \sigma(0)\frac{\omega_0}{\omega}\frac{2-be^{-b}}{1+e^{-b}}. \quad (42)$$

As one can see from (42), at $a \gg 1$, $b \gg 1$, the electrical conductivity does not depend on the magnitude of $b$ (nor $g$) and is twice as big as in the case $g \to 0$:

$$\sigma_{zz}(B) = 2\sigma(0)\frac{\omega_0}{\omega}. \quad (43)$$

In this case as well, the conductivity is monotonically decreasing as the magnetic field increases.

Now we investigate the electrical conductivity in the temperature range $k_0 T \ll \hbar\omega_0$ at relatively small magnitude of the magnetic field, satisfying the condition $b \ll 1$ ($|g|\mu_B B \ll k_0 T$). Expanding (42) into series we obtain for this case the following:

$$\sigma_{zz}(B) = \sigma(0)\left[1 - \frac{1}{2}\left(\frac{eB}{m\omega_0}\right)^2 + \left(\frac{g\mu_B B}{2k_0 T}\right)^2\right]. \qquad (44)$$

For the change of the resistance in the magnetic field from (44) we get

$$\frac{\Delta\rho(B)}{\rho(0)} = \left(\frac{\omega_c}{\omega_0}\right)^2\left[\frac{1}{2} - \left(\frac{g\hbar\omega_0}{4k_0 T}\right)^2\left(\frac{m}{m_0}\right)^2\right]. \qquad (45)$$

From (45) one can see that below certain temperature $T_0$ the magneto-resistance becomes negative. This temperature depends on the g-factor and the quantum well parameter, $\omega_0$:

$$T_0 = \frac{|g|\hbar\omega_0}{2\sqrt{2}k_0}\frac{m}{m_0}. \qquad (46)$$

For $InSb$ ($\hbar\omega_0 = 7.5\, meV$ [24]) the value is $T_0 \approx 25\, K$, and for $GaAs$ ($\hbar\omega_0 = 2.91\, meV$ [25]) the value is $T_0 \approx 0.34\, K$. Above $T_0$ the magneto-resistance is monotonically increasing in the entire range of the magnetic field.

As it can be seen from (40) and (45), at $g \to 0$ the NMR does not arise at any temperature. Therefore, the occurrence of the NMR in a quantum well is related to spin splitting of the oscillator levels.

From (45) it follows that the NMR arises under the condition $\hbar\omega_0 \gg k_0 T$, when only the lower zone of size quantization takes part in the process. Without magnetic field both spin sub-zones of the first size-quantized zone are equally populated and available for scattering. As the magnetic field increases the energy gap between spin-split sub-zones is growing, and, correspondingly, the population of the upper spin sub-zone decreases. The probability of scattering decreases (one of the channels of electronic scattering is gradually switching off) whereas the total number of the carriers remains constant. This is precisely the reason of decrease in resistance. With further growth of the magnetic field the usual mechanism of growth of resistance in the magnetic field switches on. When the cyclotron frequency exceeds the frequency that characterizes the size quantization, $\omega_0$, the magneto-resistance becomes positive. The maximal value of the NMR occurs at $\omega_c \approx 0.5\omega_0$.

The NMR arises not only in a non-degenerated case but in a weakly degenerated case as well. The range of the magnetic field in which the NMR arises without degeneration can be determined from (38), and for the case of the weak degeneration – from (37). Fig. 2 presents the plot of the relative change in the magneto-resistance with magnetic field for an *InSb* quantum well $(T = 4.2\,K,\ n = 10^9\,cm^{-2})$. As one can see from the graph, the NMR is observed in the interval of the magnetic field $(0; 1.8\,T)$. In the range $B < B_m \approx 0.5\,T$ the resistance is monotonically decreasing and at $B_m \approx 0.5\,T$ the relative change is the largest and makes about 40%. In the magnetic fields above $B_m$ the magneto-resistance is positive and is a monotonically increasing function of the magnetic field. As the temperature increases the range in which the NRM occurs is expanding whereas the relative change in resistance falls. For example, at the same **surface density** and *T=20 K*, $B_m \approx 0.8\,T$ and the maximal relative change is about 6%.

### 4.2. Strongly degenerated electronic gas

For the case of strongly degenerated electron gas, satisfying the condition (24), the expression for the conductivity can be obtained from (36). After the replacement $-\dfrac{\partial f_0}{\partial x} = \delta(x - \eta)$ and integration, taking into account (25) and (26), we get the following expression for the conductivity:

$$\sigma_{zz}(B) = \sigma(0)\frac{\omega_0}{\omega}\frac{2}{L} \qquad (47)$$

Here $\sigma(0)$ is determined by (39), and $L$ is the number of the sub-levels below the chemical potential (each level is split into two). As the magnetic field increases, starting from some value of *B* the sub-levels move upwards crossing the chemical potential one by one, with *L* changing by one unit every time. Hence, in this range of the magnetic field the electrical conductivity grows by jumps. As a result, the resistance of a quantum well with a strongly degenerated electron gas becomes a saw-tooth function of the magnetic field.

Strictly speaking, expression (47) is applicable for the case $T \to 0$. Nevertheless, this formula with reasonable accuracy can be applied for non-zero temperatures as well. Fig. 3 presents the dependence of the relative change in the magneto-resistance with magnetic field for an *InSb* quantum well at 4.2 *K* and density $n = 2.5 \cdot 10^{11}\,cm^{-2}$. The solid line (1) corresponds to the general case described by expression (37), and the dashed line (2) corresponds to the strongly degenerated case described by (47). As one can see from the graph, (47) gives a fairly accurate result, except for the jump being not as sharp.

### 5. CONCLUSION

In the present paper we studies the effect of the magnetic field on the resistance of the electron gas in a quantum well with parabolic potential when the electrons are scattered by acoustic phonons, point defects and alloy disorders, taking into account the effect of spin on the energy spectrum. The main result is that the NMR arises because of the specific energy spectrum of electrons in a quantum well with parabolic potential and because of the spin splitting of the oscillator levels for a non-degenerated electronic gas. To observe the NMR in the experiment the degeneracy must be absent. In an *InSb* quantum well, with a large *g*-factor, the NMR occurs in the temperature range $T \leq 25 K$ and in a quantum well *GaAs*, with a small *g*-factor, the NMR can arise only at very low temperatures $(T \leq 0,34 K)$.

# FIGURES

FIG. 1. The spin splitting of oscillator levels in quantum well

FIG. 2. The dependence of the relative change of the magneto-resistance with magnetic field for an $InSb$ quantum well with weakly degenerated electron gas ($T = 4.2\,K$, $n = 10^9\,cm^{-2}$).

FIG. 3. The dependence of the relative change of the magneto-resistance with magnetic field for an $InSb$ quantum well with strongly degenerated electron gas ($T = 4.2\,K$, $n = 2.5 \cdot 10^{11}\,cm^{-2}$).

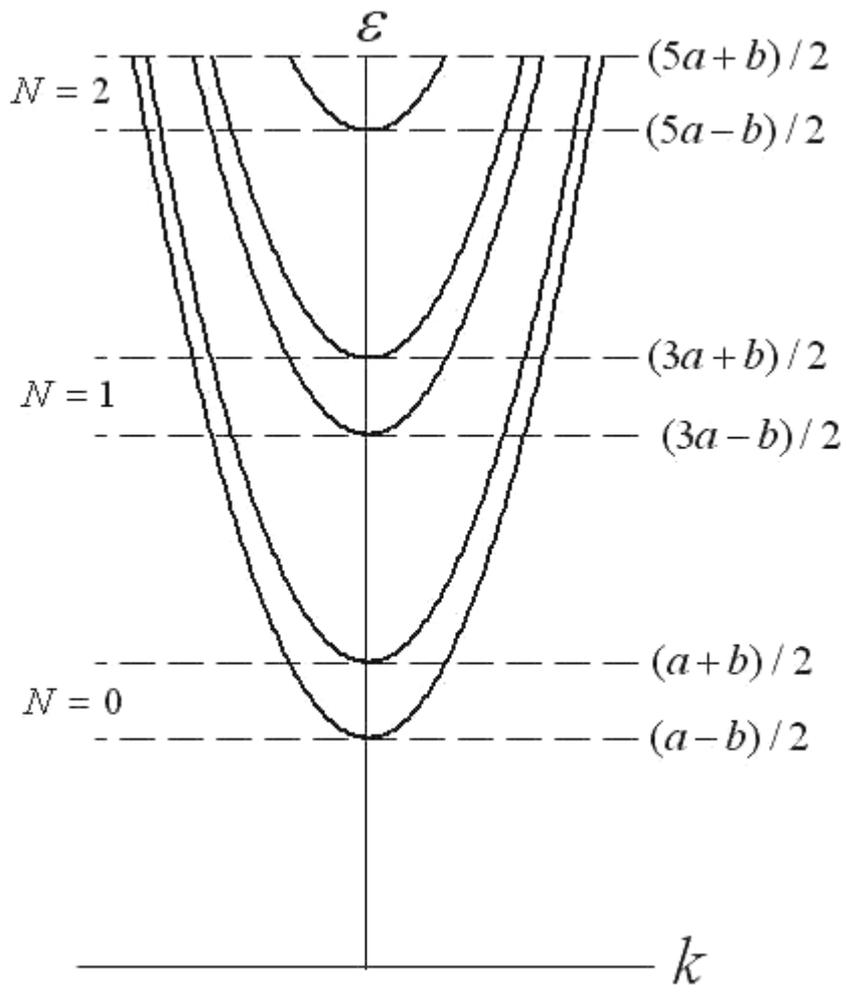

Fig. 1.

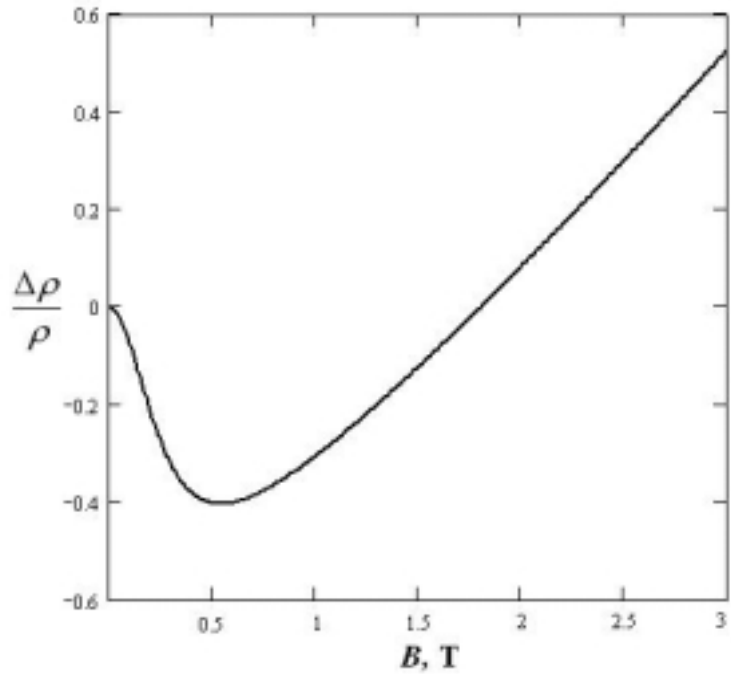

Fig. 2.

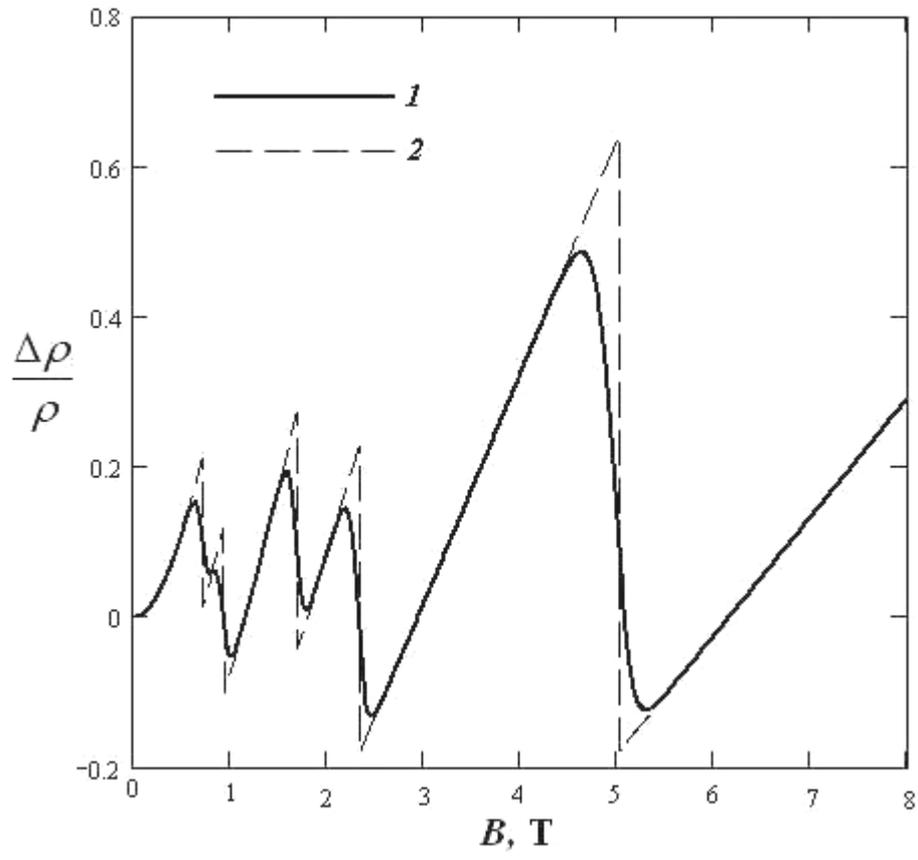

Fig. 3.